\documentclass[12pt]{article}

\begin{document}

\newcommand{\Om}{\Omega}
\newcommand{\df}{\stackrel{\rm def}{=}}
\newcommand{\co}{{\scriptstyle \circ}}
\newcommand{\de}{\delta}
\newcommand{\lb}{\lbrack}
\newcommand{\rb}{\rbrack}
\newcommand{\rn}[1]{\romannumeral #1}
\newcommand{\msc}[1]{\mbox{\scriptsize #1}}
\newcommand{\dsp}{\displaystyle}
\newcommand{\scs}[1]{{\scriptstyle #1}}

\newcommand{\ket}[1]{| #1 \rangle}
\newcommand{\bra}[1]{| #1 \langle}
\newcommand{\vac}{| \mbox{vac} \rangle }

\newcommand{\e}{\mbox{{\bf e}}}
\newcommand{\va}{\mbox{{\bf a}}}
\newcommand{\bc}{\mbox{{\bf C}}}

\newcommand{\com}{C\!\!\!\!|}

\newcommand{\br}{\mbox{{\bf R}}}
\newcommand{\bz}{\mbox{{\bf Z}}}
\newcommand{\bq}{\mbox{{\bf Q}}}
\newcommand{\bn}{\mbox{{\bf N}}}
\newcommand {\eqn}[1]{(\ref{#1})}

\newcommand{\cp}{\mbox{{\bf P}}^1}
\newcommand{\n}{\mbox{{\bf n}}}
\newcommand{\sbz}{\msc{{\bf Z}}}
\newcommand{\sn}{\msc{{\bf n}}}

\newcommand{\be}{\begin{equation}}\newcommand{\ee}{\end{equation}}
\newcommand{\bea}{\begin{eqnarray}} \newcommand{\eea}{\end{eqnarray}}
\newcommand{\ba}[1]{\begin{array}{#1}} \newcommand{\ea}{\end{array}}

\newcommand{\cleqn}{\setcounter{equation}{0}}

\makeatletter

\@addtoreset{equation}{section}

\def\theequation{\thesection.\arabic{equation}}
\makeatother

\def\np{Nucl. Phys. {\bf B}}
\def\pl{Phys. Lett. {\bf B}}
\def\mpl{Mod. Phys. {\bf A}}
\def\ijmp{Int. J. Mod. Phys. {\bf A}}
\def\cmp{Comm. Math. Phys.}
\def\prd{Phys. Rev. {\bf D}}

\def\oa{\bigcirc\!\!\!\! a}
\def\ob{\bigcirc\!\!\!\! b}

\def\vu{\vec u}
\def\vs{\vec s}
\def\vv{\vec v}
\def\vt{\vec t}
\def\vn{\vec n}
\def\ve{\vec e}
\def\vp{\vec p}
\def\vk{\vec k}
\def\vx{\vec x}
\def\vz{\vec z}
\def\vy{\vec y}


\begin{flushright}
La Plata Th/02-01\\March, 2002
\end{flushright}

\bigskip

\begin{center}

{\Large\bf On supersymmetric $Dp$-$\bar D p$ brane solutions} \footnote{ This work was
partially supported by CONICET, Argentina}

\bigskip
\bigskip

{\it \large Adri\'{a}n R. Lugo} \footnote{ {\sf lugo@obelix.fisica.unlp.edu.ar}
}
\bigskip

{\it Departamento de F\'\i sica, Facultad de Ciencias Exactas \\
Universidad Nacional de La Plata\\ C.C. 67, (1900) La Plata,
Argentina}
\bigskip
\bigskip

\end{center}

\begin{abstract}
We analyze in the spirit of hep-th/0110039 the possible existence of supersymmetric $D
p$-$\bar D p$ brane systems in flat ten dimensional Minkowski space. For $p=3,4$ we
show that besides the solutions related by T-duality to the $D2$-$\bar D 2$ systems
found by Bak and Karch there exist other ansatz whose compatibility is shown from
general arguments and that preserve also eight supercharges, in particular a $D4$-$\bar
D4$ system with $D2$-branes dissolved on it and Taub-NUT charge. We carry out the
explicit construction in Weyl basis of the corresponding Killing spinors and conjecture
the existence of new solutions for higher dimensional branes with some compact
directions analogous to the supertube recently found.

\end{abstract}

\bigskip

\section{Introduction }
\cleqn

The discovery in type II string theories of cylinder-like branes preserving a quarter of the supersymmetries of
the flat Minkowski space-time, the so-called ``supertubes" \cite{sutubo1}, \cite{sutubo2}, \cite{cho} has attracted
much attention recently.
The stabilizing factor at the origin of their BPS character that prevent them from
collapse is the angular momentum generated by the non-zero gauge field living on the
brane.
The solution in  \cite{sutubo1}  presenting circular section was extended to arbitrary section
in \cite{sutubo3}; supertubes in the matrix model context can be found in \cite{baklee}, \cite{bakkim}.

A feature of the supertube is that it has $D 0$ and $F1$ charges, but not $D2$ charge.
An interesting observation related to this fact was made by Bak and Karch (BK); if we take the elliptical supertube
with semi-axis $a$ and $b$ in the limit when for example $a$ goes to infinity that is equivalent to see the geometry
near the tube where it looks flat, the system should become like two flat branes separated by a distance $b$.
But because of the absence of $D2$ charge it is natural to suspect that indeed the
system could be a $D2$-$\bar D 2$ one.
The existence of this system as well as systems with arbitrary numbers of $D2$ and $\bar D 2$ branes was proved in the
context of the Born-Infeld action in reference \cite{bakar} where the conditions to be satisfied by the Killing spinors
were identified, while the absence of tachyonic instabilities was shown in \cite{sutubo2}, \cite{bakota}.
The aim of this letter is to extend the results of \cite{bakar} to higher dimensional brane-antibrane systems.
In the course of the investigation we will find, other than the T-dual solutions to that of BK, also new solutions
for $p=3,4$.

We start by remembering   some relevant facts.
Let $\{ X^M(\xi), M=0,1,\dots,9 \}$ the embedding fields in a ten-dimensional space-time of a
$D p$ -brane parameterized by coordinates
$\{\xi^\mu, \mu=0,1,\dots,p\}$ and $\{A_\mu(\xi), \mu=0,1,\dots,p\}$ the abelian gauge field
living on it, $F=dA$ being the field strength.
Let $\epsilon$ be a general Killing spinor of some background
$(G_{MN}, B_{MN}, \phi, A^{(p+1)}_{M_1\dots M_{p+1}})$
of type II string theory.
Then the introduction of the brane in such space will preserve the supersymmetries that
satisfy \cite{berg2}
\be
\Gamma\;\epsilon = +\;\epsilon \label{susycond}
\ee
where the $\Gamma$-matrix is
defined by \cite{berg1}
\bea
\Gamma &\equiv& \frac{1}{ |d|^\frac{1}{2}  }\;
\sum_{n=0}^{\left[\frac{p+1}{2}\right]}\; \frac{1}{2^n\,n!}\; f_{\mu_1\nu_1}\dots
f_{\mu_n\nu_n}\;\gamma^{\mu_1\nu_1\dots\mu_n\nu_n}\; \left\{ \begin{array}{ll}
(\Gamma_{11})^{n + \frac{p-2}{2}}\;\Gamma_{(0)} &,\;IIA\cr
(-)^n\;(\sigma_3)^{n+\frac{p-3}{2}}\;i\,\sigma_2\otimes\Gamma_{(0)}&,\;IIB\end{array}\right.\cr
\Gamma_{(0)} &\equiv&
\frac{1}{(p+1)!}\;\epsilon_{\mu_1\dots\mu_{p+1}}\;\gamma^{\mu_1\dots\mu_{p+1}}
\label{gama}
\eea
In all these expressions the pull-back of the background fields to the brane defined by
\be
t_{\mu_1\dots\mu_n}(\xi) \equiv T_{M_1\dots M_n}(X)|_{X(\xi)}\;
\partial_{\mu_1} X^{M_1}(\xi)\dots\partial_{\mu_n} X^{M_n}(\xi)
\ee
for any tensor field $T_{M_1\dots M_n}$ as well as $\gamma_\mu \equiv \bar E_\mu{}^A\, \Gamma_A$
are understood, where $\bar E_\mu{}^A\equiv E_M{}^{A}(X)|_{X(\xi)}\, \partial_\mu X^M(\xi)$
is the ``pull-backed" vielbein defined by
$g_{\mu\nu}(\xi)= \eta_{AB}\,\bar E_\mu{}^A(\xi)\, \bar E_\nu{}^B(\xi)$,
$\Gamma_A$ are the flat, tangent space $\Gamma$-matrices in ten dimensions,
$d\equiv \det(\delta^\mu{}_\nu + {\it f}^\mu{}_\nu)$ and
$f^\mu{}_\nu\equiv g^{\mu\rho}\, (F_{\rho\nu} + b_{\rho\nu}) =
g^{\mu\rho}\, f_{\rho\nu}$
(we take $2\,\pi\,\alpha' =1 $) .
The induced volume form on the brane is $ \epsilon_{\mu_1\dots\mu_{p+1}} \equiv \sqrt{|g|}\,
\varepsilon_{\mu_1\dots\mu_{p+1}}$,
where $\varepsilon_{01\dots p}= +1$ in some patch defines an orientation;
condition (\ref{susycond}) with a $``-"$ sign on the rhs corresponds to
the anti-brane with the {\it same} fields of the brane, since by definition they have opposite
orientations.
Lastly, in the case of type IIA string theory the spinors must be Majorana, while in the
type IIB case we take a pair of Majorana-Weyl spinors with the same chirality (and
$\gamma^{\mu_1\nu_1\dots\mu_n\nu_n}\rightarrow 1_2\otimes\gamma^{\mu_1\nu_1\dots\mu_n\nu_n}$
should be understood in (\ref{gama})).

We are now ready to start with the analysis of various cases.
We will restrict in this paper to work on the flat ten-dimensional Minkowski vacuum
of type II string theories, the spinors being constants (in cartesian coordinates) of the type mentioned above.

\section{The $D2$-$\bar D2$ system}
\cleqn

\subsection{The Bak-Karch ansatz}

Let us consider a flat D2-brane extended in directions $(X^0,X^1,X^2)$, with field
strength $F_{20}= E , F_{12}=B$.
The general case with $F_{10}\neq 0$ can be reached from this using the Lorentz
invariance $SO(1,2)$ of the setting by means of a rotation.
We can also take $E<0$ or $E>0$, it will not be relevant to fix a particular sign.
The $\Gamma$-matrix is
\be
\Gamma = |1-E^2 + B^2|^{-\frac{1}{2}}\;\left( \Gamma_{012}
+ E\;\Gamma_{1}\;\Gamma_{11} + B\;\Gamma_{0}\;\Gamma_{11}\right)\label{gama2}
\ee

It is well-known that equation (\ref{susycond}) has solutions preserving $\frac{1}{2}$ of
SUSY, i.e. $16$ supercharges, for any constant $F_{\mu\nu}$ \cite{berg2}.
The ansatz introduced in \cite{bakar}  consists in subdividing condition
(\ref{susycond}) in two (or maybe more, if possible) parts in such a way they result
compatible; in other words they show we can get novel non-trivial solutions at expenses of
SUSY.
In particular their solutions, as well as all the solutions presented in this paper,
preserve $\frac{1}{4}$ SUSY.
We will refer to the two conditions to solve (\ref{susycond}) with the labels
${\oa}$ and ${\ob}\,$; thus we write
\bea
\Gamma_{\oa}\;\epsilon &=& \epsilon\;\;\;\;\;,\;\;\;\; \Gamma_{\oa} =
E\; \Gamma_{02}\;\Gamma_{11}\label{conda}\\
\Gamma_{\ob}\;\epsilon &=& \epsilon\;\;\;\;\;,\;\;\;\;
\Gamma_{\ob} = sg(B)\; \Gamma_{0}\;\Gamma_{11}\label{condb}
\eea
where $``sg"$ stands for the sign-function.
Equation (\ref{conda}) is equivalent to ask for the annihilation of the first two terms in
(\ref{gama2}) when acting on the spinor while (\ref{condb}) enforces (\ref{susycond}) (with a minus sign on the
rhs  when the $\bar D$-brane case is considered).
They sign out the presence of dissolved $F1$ in $\check e_2$-direction and $D0$ branes respectively \cite{bakar}.
Consistency conditions for ${\oa}$ and ${\ob}\,$ say that
\be
\Gamma_{\oa}{}^2\;\epsilon = \Gamma_{\ob}{}^2 \;\epsilon = \epsilon
\ee
But $\Gamma_{\oa}{}^2 = E^2\, 1$, so we must take $E^2=1$ for the electric field,
constraint used in writing (\ref{condb}).
Assumed it,  we get $\Gamma_{\ob}{}^2 = 1$ and so nothing new is added.
Finally the compatibility between the two conditions is assured from the fact that
$[\Gamma_{\oa}\,; \Gamma_{\ob}\,] = 0$.
The analysis of these consistency conditions will determine the compatibility conditions and will be
the route to be followed to assure the existence of such solutions.
From the further properties
$tr\Gamma_{\oa} = tr\Gamma_{\ob}= tr\Gamma_{\oa}\,\Gamma_{\ob}=0$ is straightforward to conclude that
each condition preserves $\frac{1}{2}$ SUSY and both together $\frac{1}{4}$ SUSY.

\subsection{Explicit solution}

We can obtain the Killing spinors explicitly working in the Weyl basis, we refer the
reader to the appendix for a brief review.

The relevant operators are
\be
\Gamma_{\oa} = E\; \sigma_2\,1\,1\,1\,\sigma_1 \;\;\;\;\;,\;\;\;
\Gamma_{\ob}= sg(B)\; 1\;1\;1\;1\;\sigma_1
\ee
The spinorial space is divided in two $16$-dimensional (complex) subspaces with
$\Gamma_{\oa} = \pm1$; each one is expanded by the vectors
\bea
\epsilon^{(\pm)}_{(s_1\dots s_4)} &=& (0s_1\dots s_4) \mp i\, E\; (1s_1s_2s_3\, \bar s_4)\cr
\Gamma_{\oa}\,\epsilon^{(\pm)}_{(s_1\dots s_4)} &=& \pm\;\epsilon^{(\pm)}_{(s_1\dots s_4)}\cr
\Gamma_{\ob}\,\epsilon^{(\pm)}_{(s_1\dots s_4)} &=& sg(B)\;\epsilon^{(\pm)}_{(s_1s_2s_3\,\bar s_4)}
\eea
The last property is a result of the strict commutation of both operators that allows
for the common diagonalisation in each subspace separately.
We are of course interested in the subspace with $\Gamma_{\oa} = +1$; on it we can
introduce the following set of basis vectors
\bea
\eta^{(\pm)}_{(s_1s_2s_3)} &=& \epsilon^{(+)}_{(s_1s_2s_30)} \pm sg(B)\;
\epsilon^{(+)}_{(s_1s_2s_3 1)}\cr
\Gamma_{\ob}\;\eta^{(\pm)}_{(s_1s_2s_3)} &=& \pm\;\eta^{(\pm)}_{(s_1s_2s_3)}
\eea
From here we conclude that the spinors $\{\eta^{(+)}_{(s_1s_2s_3)}\}$ ($\{\eta^{(-)}_{(s_1s_2s_3)}\}$ )
expand the eight-dimensional complex space of Killing vectors for a $D2$ ($\bar D2$) -brane
extended in $(012)$ directions with field strength defined by $(E_2=E,B), |E|=1,$ and
satisfying conditions (\ref{conda}),(\ref{condb}).
Then a general Killing spinor of the $D2$-brane admits the expansion
\be
\epsilon = \sum_{s_1,s_2,s_3}\; \epsilon^{(s_1s_2s_3)}\; \eta^{(+)}_{(s_1s_2s_3)}
\;\;\;\;,\;\;\;\;\epsilon^{(s_1s_2s_3)}\in \com\label{kill2}
\ee
The observation made by BK is that a $\bar D 2$ with {\it different} fields
$(E_2=E,-B), |E|=1,$ must have the same Killing spinors.
This fact follows immediately from the equality
$\eta^{(-)}_{(s_1s_2s_3)}|_{-B} = \eta^{(+)}_{(s_1s_2s_3)}|_B$.
Therefore we should be able to put arbitrary (parallel) number of $D2$-branes and
$\bar D 2$-branes of the characteristics defined above and such configurations must
preserve $\frac{1}{4}$ SUSY with the corresponding Killing spinors given by (\ref{kill2}).

Finally it is worth to spend some words about the Majorana condition to be imposed.
In a Majorana basis where all the $\Gamma$-matrices are real (or purely imaginary) the
constraint is straightforward because we can take $D=1$ in such a basis.
But it is not so in a Weyl basis; this is a price to be paid for working in a setting
where computations are relatively easy to handle in any dimension.
We get from imposing (\ref{majo}) on the spinor (\ref{kill2})
\be
{\epsilon^{(s_1s_2s_3)}}^* = i\;sg(B\,E)\; (-)^{1+s_1+s_3}\;
\epsilon^{(\bar s_1\,\bar s_2\,\bar s_3 )}\label{majo2}
\ee
This completes the characterization of the Killing spinors of the BK solution.

After sketched with this known example the route to follow we move to higher dimensional cases.

\section{The $D3$-$\bar D3$ system}
\cleqn

Let us consider  a flat D3-brane extended in directions $(X^0,X^1,X^2,X^3)$ with field strength
\be
(f^\mu{}_\nu) = \left(\matrix{ 0 & E_1 & E_2 & 0\cr
                             E_1 & 0   &  0  & 0\cr
                             E_2 & 0   &  0  & B\cr
                              0  & 0   & -B  & 0\cr}\right)\label{fd3}
\ee
where we have taken with no lost of generality the electromagnetic fields in the plane $(12)$.
The $\Gamma$-matrix is then given by
\bea
\Gamma &=& |d|^{-\frac{1}{2}} \;\big( i\,\sigma_2\otimes \Gamma_{0123} -
\sigma_1\otimes(E_1\,\Gamma_{23}+ E_2\,\Gamma_{31} - B\,\Gamma_{01})- E_1\;B\;
i\,\sigma_2\otimes 1_{32}\big)\cr
d &=& 1- E_1{}^2 - E_2{}^2 + B^2 - E_1{}^2\,B^2
\label{gama3}
\eea
The equation to solve (\ref{susycond}) is written as
\be
\Gamma\; \left( \begin{array}{l} \epsilon_1 \cr\epsilon_2 \end{array}\right) =
\left( \begin{array}{l} \epsilon_1 \cr\epsilon_2 \end{array}\right)\label{susycond3}
\ee
where $(\epsilon_1, \epsilon_2)$ are Majorana-Weyl spinors of the same chirality.
A minus sign on the rhs applies for the antibrane.
We have found two possible solutions to (\ref{susycond3}).

\subsection{Solution I}

Consistency of ${\oa}$  will require the constraint $E_1{}^2+E_2{}^2=1$ which implies
$d=B^2\,E_2{}^2$ and then $B$ and $E_2$ non zero, fact that we will assume; the ansatz is
\bea
\Gamma_{\oa}\;\epsilon_1 &=& \epsilon_1 \;\;\;\;\;,\;\;\;\;,
\Gamma_{\oa}\;\epsilon_2 = -\epsilon_2\;\;\;\;\;\;\;\;\;\;;\;\;\;\;\;\;
\Gamma_{\oa} = -E_1\; \Gamma_{01}-E_2\; \Gamma_{02}\label{conda31}\\
\Gamma_{\ob}\;\epsilon_1 &=& \epsilon_2\;\;\;\;\;,\;\;\;\; \Gamma_{\ob} = \frac{sg(B)}{|E_2|}\;(E_1\,1 +
\Gamma_{01})
\label{condb31}
\eea
with a minus sign in (\ref{condb31}) for the $\bar D 3$-brane.
The compatibility for both constraints requires that
\be
\Gamma_{\oa}\,\Gamma_{\ob}\,\Gamma_{\oa}{}^{-1}\,\epsilon_1 = - \Gamma_{\ob}\,\epsilon_1\label{compab3}
\ee
A short computation shows
\be
\{\Gamma_{\oa}\,; \Gamma_{\ob}\} = \frac{2\,E_1\,sg(B)}{|E_2|}\; (\Gamma_{\oa}-1)
\ee
that says that in the subspace with $\Gamma_{\oa}=1$ to which $\epsilon_1$ belongs
(\ref{compab3}) is obeyed.

The explicit solution can be worked out; first we introduce ( $E_\pm = E_1 \pm i\,E_2$)
\bea
\epsilon^{(\pm)}_{(s_1\dots s_4)} &=& (0s_1\dots s_4) \mp (-)^{\sum_{k=1}^4\,s_k}\;
E_+\; (1s_1s_2s_3\, \bar s_4)\cr
\Gamma_{\oa}\,\epsilon^{(\pm)}_{(s_1\dots s_4)} &=& \pm\;\epsilon^{(\pm)}_{(s_1\dots s_4)}\cr
\Gamma_{\ob}\,\epsilon^{(+)}_{(s_1\dots s_4)} &=& -i\,sg(B\,E_2)\;
\epsilon^{(-)}_{(s_1\dots s_4)}
\eea
The last line shows the compatibility of both constraints; the imposition of (\ref{condb31})
gives for both spinors the following general form
\bea
\epsilon_1 &=& \sum_{s_1,\dots,s_4}\; \epsilon^{(s_1\dots s_4)}\; \epsilon^{(+)}_{(s_1\dots s_4)}\cr
\epsilon_2 &=& -i\,sg(B\,E_2)\;\sum_{s_1,\dots,s_4}\; \epsilon^{(s_1\dots s_4)}\; \epsilon^{(-)}_{(s_1\dots s_4)}
\label{kill31}
\eea
The complex parameters $ \epsilon^{(s_1\dots s_4)}$ expand a $32$-dimensional
space; however the spinors must be Weyl and Majorana.
Because $\Gamma_{11}(\epsilon^{(+)}_{(s_1\dots s_4)}) = (-)^{1+\sum_{k=1}^4\,s_k}$ we must
constraint this value to $+1\,(-1)$ if we decide to take both of them left- (right) handed,
so one index, e.g. $s_4$, will be fixed by this condition.
Finally the Majorana condition
\be
{\epsilon^{(s_1\dots s_4)}}^* = E_+\;(-)^{s_2+s_4}\; \epsilon^{(\bar s_1\dots \bar s_4)}
\label{majo31}
\ee
shows that the solution has indeed $8$ supercharges.
Conditions (\ref{conda31}), (\ref{condb31}) can be interpreted as dissolved fundamental strings in the plane
$(12)$ at angle $\arctan \frac{E_2}{E_1}$ and $D1$-branes in the $\check e_1$-direction respectively.
This solution for a $D3$-brane can be extended to a $\bar D 3$-brane with the same Killing
spinors by reverting the direction of the magnetic field, $B\rightarrow -B$, as in the BK solution.
In fact it is easy to see that we can reach it by T-dualizing in $\check e_1$ and boosting with $\beta=-E_1$ in
that direction, remaining with the dual $D2$- brane in the $(023)$- hyperplane, electric field
$E= sg(E_2)$ in $\check e_2$ and magnetic field $B$.

\subsection{Solution II}

Here we will work out a new ansatz.
The idea is to ask for the cancellation of the field-dependent part in (\ref{gama3}).
Compatibility of this ansatz imposes the constraints $E_1=0$ and $B^2 = E_2{}^2\neq 0$,
i.e. it is just a solution for orthogonal electric and magnetic fields with the same module.
The corresponding conditions are
\bea
\Gamma_{\oa}\;\epsilon_i &=& \epsilon_i \;,i=1,2\;\;\;,\;\;\;
\Gamma_{\oa} = -sg(B\, E_2)\; \Gamma_{03}\label{conda32}\\
\Gamma_{\ob}\;\epsilon_1 &=& \epsilon_2\;\;\;\;\;,\;\;\;\; \Gamma_{\ob} = -\Gamma_{0123}
\label{condb32}
\eea
The further properties $\Gamma_{\oa}{}^2 = -\Gamma_{\ob}{}^2 =1\;,\;
[\Gamma_{\oa}\,;\Gamma_{\ob}]=0$
assure the existence of solutions to these equations.

To get the Killing spinors associated to these systems we first solve ${\oa}$ through
\bea
\epsilon^{(+)}_{(s_1\dots s_4)} &=& (s_1\dots s_40) + sg(B\, E_2)\;(-)^{1+\sum_{k=2}^4\,s_k}\;
(s_1\, \bar s_2\,s_3s_4 1)\cr
\Gamma_{\oa}\,\epsilon^{(+)}_{(s_1\dots s_4)} &=& \epsilon^{(+)}_{(s_1\dots s_4)}\cr
\Gamma_{\ob}\,\epsilon^{(+)}_{(s_1\dots s_4)} &=& i\,sg(B\,E_2)\; (-)^{s_1}\;
\epsilon^{(+)}_{(s_1\dots s_4)}
\eea
Then (\ref{condb32}) yields the general solution
\bea
\epsilon_1 &=& \sum_{s_1,\dots,s_4}\; \epsilon^{(s_1\dots s_4)}\;
\epsilon^{(+)}_{(s_1\dots s_4)}\cr
\epsilon_2 &=& i\,sg(B\,E_2)\;\sum_{s_1,\dots,s_4}\;(-)^{s_1}\; \epsilon^{(s_1\dots s_4)}\;
\epsilon^{(+)}_{(s_1\dots s_4)}
\label{kill32}
\eea
As with the Solution I a Weyl left (or right) condition
$(-)^{1+\sum_{k=1}^4\,s_k} = +1(-1)$
must be imposed; together with the Majorana constraint
\be
{\epsilon^{(s_1\dots s_4)}}^* = (-)^{1+s_2+s_4}\; \epsilon^{(\bar s_1\dots \bar s_4)}
\label{majo32}
\ee
it is shown that the solution preserves $\frac{1}{4}$ SUSY.
Furthermore conditions (\ref{conda32}) and (\ref{condb32}) correspond to the SUSY's preserved by a p-p wave
moving on $\check e_3$ direction and a $D3$-brane in $(0123)$, the existence of such configuration of
intersecting branes being known since time ago (see for example reference \cite{gaun}); the SUSY of the
$D3$-brane configuration results then only sensitive to that of the constituents induced by the background fields.

On the other hand for the $\bar D3$-branes (\ref{condb32}) has a minus sign on the rhs;
what is the same it is replaced by $\Gamma_{\ob}\;\epsilon_2 = \epsilon_1$, i.e. both
spinors are interchanged, operation that obviously leaves invariant (\ref{conda32}).
In view of the existence of the $O(2)$ automorphism group in type IIB string theory
which rotates the supersymmetry generators we are tempted to conjecture the existence of
systems of branes and antibranes in arbitrary number provided that on each one live fields
$(E_2, B)$ with $|E_2|=|B|$ but otherwise arbitrary,
the only condition common to all of them being to have the same $sg(B\,E_2)$.

\section{The $D4$-$\bar D4$ system}
\cleqn

Now we take a flat D4-brane extended in directions $(X^0,X^1,X^2,X^3,X^4)$
\be
(f^\mu{}_\nu) = \left(\matrix{ 0  & E_1 &  0  & E_2 & 0\cr
                              E_1 & 0   & B_1 &  0   & 0\cr
                               0  &-B_1  & 0  &  0   & 0\cr
                              E_2 & 0   &  0  &  0   & B_2\cr
                               0  & 0   &  0  & -B_2 & 0 \cr}\right)\label{fd4}
\ee
We again do not loose generality doing so because the anti-symmetric matrix
$(f^i{}_j)= (F_{ij}), i,j=1,\dots,4\,,$ can be put in the standard form
$i\sigma_2\otimes\left(\matrix{B_1&0\cr 0&B_2}\right)$
by means of a $SO(4)$-rotation; a $SO(2)\times SO(2)$ is then left over that we fix by
putting the electric field $E_i\equiv F_{i0}$ to the form in (\ref{fd4})
\footnote{
More generically, any anti-symmetric matrix in $d$-dimensions can be written as
$B= k^t\, d \, k$ where $k\in SO(d)/SO(2)^{[\frac{d}{2}]}$ and
$d = i\sigma_2\otimes diag(B_1,\dots,B_{[\frac{d}{2}]})$
(a $0$ is added if $d$ is odd).
It can be shown that in the general case when the magnetic field matrix $(B_{ij})\equiv (F_{ij})$
has inverse there is no boost compatible with (\ref{fd4}).
However it is possible in this case to go to a frame where there is no electric field, being the boost
velocity  $\vec\beta = - B^{-1}\,\vec E$,  if the condition $\beta^2= {\vec E}^t\, (-B^2)^{-1}\,\vec E<1$
is fulfilled.
}.
The $\Gamma$-matrix is
\bea
\Gamma &=& |d|^{-\frac{1}{2}} \;\left( \Gamma_{11}\;\Gamma_{01234} +
E_1\,\Gamma_{234}+ E_2\,\Gamma_{124} - B_1\,\Gamma_{034}-B_2\,\Gamma_{012}\right.\cr
&+& \left. E_1\;B_2\; \Gamma_2\;\Gamma_{11} + E_2\;B_1\; \Gamma_4\;\Gamma_{11}
- B_1\;B_2\; \Gamma_0\;\Gamma_{11}\right)\cr
d &=& 1- E_1{}^2 - E_2{}^2 + B_1{}^2 + B_2{}^2 + B_1{}^2\;  B_2{}^2-
E_1{}^2\; B_2{}^2 - E_2{}^2\; B_1{}^2\label{gama4}
\eea
We have found two solutions to equation (\ref{susycond}).

\subsection{Solution I}

Consistency of ${\oa}$ will require the constraint
$E_1{}^2+E_2{}^2=1$ which we will assume; the ansatz is as in (\ref{conda}),
(\ref{condb})
where the relevant operators are
\bea
\Gamma_{\oa} &=& (E_1\; \Gamma_{01} + E_2\; \Gamma_{03})\;\Gamma_{11}\label{conda41}\\
\Gamma_{\ob} &=& d^{-\frac{1}{2}}\;
\big(- B_1\,\Gamma_{034} -B_2\,\Gamma_{012} + (E_1\;B_2\; \Gamma_2 +
E_2\;B_1\; \Gamma_4 - B_1\;B_2\; \Gamma_0)\Gamma_{11} \big)\label{condb41}
\eea
where $d = B_1{}^2\, B_2{}^2 +  B_1{}^2\, E_1{}^2 + B_2{}^2\, E_2{}^2>0$.
Assuming ${\oa}$ is obeyed, the consistency conditions
$(\Gamma_{\ob}{}^2 -1)\epsilon = [\Gamma_{\oa}\,;\Gamma_{\ob}\,]\,\epsilon = 0$
follow from
\bea
\Gamma_{\ob}{}^2 &=& 1 + \frac{2\,B_1\,B_2}{d}\;\Gamma_{1234}\;(\Gamma_{\oa}-1)\cr
[\Gamma_{\oa}\,; \Gamma_{\ob}\,] &=& 2\,(E_1\;B_2\;\Gamma_2 + E_2\;B_1\;\Gamma_4)\;\Gamma_{11}\;
(1-\Gamma_{\oa})
\eea
From standard arguments we should get again a solution preserving $\frac{1}{4}$ SUSY.

The explicit solution can be obtained as it was made in the precedent cases; first we introduce
a basis for the subspace of spinors obeying ${\oa}\,$,
\bea
\epsilon^{(+)}_{(s_1\dots s_4)} &=& (s_1\dots s_40) - E_1\; (\bar s_1\,s_2s_3s_41)
- E_2\; (-)^{s_1}\; (s_1\,\bar s_2 s_3 s_4 1)\cr
\Gamma_{\oa}\;\epsilon^{(+)}_{(s_1\dots s_4)} &=& +\;\epsilon^{(+)}_{(s_1\dots s_4)}
\eea
In terms of them we can express a basis which also diagonalice $\Gamma_{\ob}\,$,
\bea
\eta^{(\pm)}_{(s_1s_2s_3)} &=& \epsilon^{(+)}_{(\bar s_1\,s_1 s_2 s_3)} \pm a(s)\;
\epsilon^{(+)}_{(s_1 s_1 s_2 s_3)} \pm b(s)\; \epsilon^{(+)}_{(\bar s_1\,\bar s_1\, s_2
s_3)}\cr \Gamma_{\ob}\;\eta^{(\pm)}_{(s_1s_2s_3)} &=& \pm\;\eta^{(\pm)}_{(s_1s_2s_3)}
\label{kill41} \eea where \bea a(s) &=& \frac{E_1\,B_1}{\sqrt{d}}\;\left( B_2 +
i\,(-)^{s_1+s_2+s_3}\right)\cr b(s) &=& \frac{E_2\,B_2}{\sqrt{d}}\;\left(
(-)^{1+s_1}\,B_1 + i\,(-)^{s_2+s_3}\right) \label{kill41bis} \eea So we conclude that
$\{\eta^{(+)}_{(s_1s_2s_3)}\}$ is a basis of Killing spinors for (parallel
superposition of) $D4$-branes, while $\{\eta^{(-)}_{(s_1s_2s_3)}\}$ it is for a $\bar D
4$-brane (with the same fields as the $D4$-brane).
However to get brane-antibrane BPS systems the preserved supersymmetries must coincide.
Condition ${\oa}$ implies that both $D4$ and $\bar D 4$ branes must have the same electric
field, but ${\ob}$ implies that the magnetic fields must have opposite signs and not only
this, direct inspection of (\ref{condb41}) (or (\ref{kill41})) shows that $B_1\,B_2=0$
necessary must hold.
This gives two possible solutions, related by permutations of the planes $(12)$ and $(34)$.

\bigskip

\noindent\underline{ Case $B_1=0$; $d=B_2{}^2\,E_2{}^2$}

Then $E_2$ and $B_2$ are non zero; the $\bar D4$-brane will have $E_2$ and $-B_2$ fields
as said.
From (\ref{kill41}), (\ref{kill41bis}) the Killing spinors reduce to
\be
\eta^{(+)}_{(s_1s_2s_3)} = \epsilon^{(+)}_{(\bar s_1\,s_1 s_2 s_3)} +
i\, sg(E_2\,B_2)\; (-)^{s_2+s_3}\;\epsilon^{(+)}_{(\bar s_1\,\bar s_1\, s_2 s_3)}
\label{kill411}
\ee

\noindent\underline{Case $B_2=0$; $d=B_1{}^2\,E_1{}^2$}

Now $E_1$ and $B_1$ must be non zero; the $\bar D4$-brane will have $E_1$ and $-B_1$ fields.
The Killing spinors reduce to
\be
\eta^{(+)}_{(s_1s_2s_3)} = \epsilon^{(+)}_{(\bar s_1\,s_1 s_2 s_3)} +
i\, sg(E_1\,B_1)\; (-)^{s_1+s_2+s_3}\;\epsilon^{(+)}_{(s_1s_1s_2 s_3)}\label{kill412}
\ee

Again we can identify (\ref{conda41}) and (\ref{condb41})  with $F1$ charge in $(01)$
and $D2$-brane charge in $(034)$ respectively; these solutions are T-duals to the
Solutions I just considered (for example in the last case, by means of a T-duality in
$\check e_3$, a boost in that direction with $\beta= - E_2$ and further T-duality in
$\check e_4$ we go back to BK solution).
The Majorana condition for a generic Killing spinor (\ref{kill2}) reads
\be
{\epsilon^{(s_1s_2s_3)}}^* = (-)^{1+s_1+s_3}\;\epsilon^{(\bar s_1\bar s_2\bar s_3)}
\label{majo41}
\ee

\subsection{Solution II}

Here we present a new solution giving a $D4$-$\bar D 4$ system.
The ansatz ${\oa}$ this time consists in imposing the cancellation of the part of
(\ref{gama4}) even in the fields by itself.
Consistency imposes the constraint
\be
B_1{}^2\,B_2{}^2 - B_1{}^2\, E_2{}^2 - B_2{}^2\,E_1{}^2 = 1 \label{const42}
\ee
which replaces the ${\vec E}^2 = 1$ constraint of the BK-type solutions and that we
will assume henceforth.
Let us note that $|B_1\,B_2|\geq 1$, in particular $B_{ij}$ must be non singular;
also from (\ref{const42}) $d=2 - E_1{}^2 - E_2{}^2 + B_1{}^2 + B_2{}^2 \geq 2$.

The corresponding conditions are as in (\ref{conda}), (\ref{condb})
where the relevant operators are
\bea
\Gamma_{\oa} &=& -B_1\, E_2\; \Gamma_{0123} - B_2\, E_1\;\Gamma_{0134} - B_1\, B_2\;\Gamma_{1234}
\label{conda42}\\
\Gamma_{\ob} &=& d^{-\frac{1}{2}}\;
\left( E_1\,\Gamma_{234} + E_2\,\Gamma_{124} - B_1\; \Gamma_{034} - B_2\;\Gamma_{012}\right)\label{condb42}
\eea
Consistency of the ansatz follows from $\;
\Gamma_{\ob}{}^2 = 1 - \frac{1}{d}\; (\Gamma_{\oa}-1)^2\;\;,\;\;
[\Gamma_{\oa}\, ;\Gamma_{\ob}\,] = 0\;$,
which shows that there should exists solution preserving eight supersymmetries.
In order to write it we introduce a basis in which $\Gamma_{\oa}$ is diagonal
\bea
\tilde\epsilon_{(s_1\dots s_5)} &=& (s_1\dots s_5) - i\,\alpha_1\; (-)^{\sum_{k=2}^{5}}\;
(\bar s_1 s_2s_3s_4 \bar s_5)\cr
&-& i\,\alpha_2\; (-)^{\sum_{k=3}^{5}}\; (s_1\bar s_2\, s_3s_4\bar s_5)\cr
\Gamma_{\oa}\;\tilde\epsilon_{(s_1\dots s_5)} &=& sg(B_1\,B_2)\,(-)^{s_1+s_2}\;
\tilde\epsilon_{(s_1\dots s_5)}
\eea
where
$\alpha_1 = \frac{sg(B_1\,B_2)\,B_2\, E_1}{1 + |B_1\,B_2|}\;,\;
 \alpha_2 = \frac{sg(B_1\,B_2)\,B_1\, E_2}{1 + |B_1\,B_2|} $.
From here it is clear that a basis for the space obeying $\Gamma_{\oa} \,\epsilon = \epsilon$
consists of the spinors
\be
\epsilon^{(+)}_{(s_2s_3s_4s_5)} \equiv \tilde\epsilon_{(s_1\dots s_5)} |_{ (-)^{s_1} =
sg(B_1 B_2) (-)^{s_2}  }
\ee
With the definitions
\bea
A_1 &=& \frac{B_1\,(1+B_2{}^2 - E_2{}^2) + sg(B_1\,B_2)\;B_2\,(1+B_1{}^2 - E_1{}^2) }
{\sqrt{d}\;(1+ |B_1\,B_2|)}\cr
A_2 &=& \frac{E_1\,E_2\;(B_2 - sg(B_1\,B_2)\; B_1)}{\sqrt{d}\;(1+ |B_1\,B_2|)}
\eea
($(A_1{}^2 + A_2{}^2 = 1)$) the Killing spinors result
\bea
\eta^{(\pm)}_{(s_1s_2s_3)} &=& \epsilon^{(+)}_{(s_1s_2s_3 1)} \mp i\, sg(B_1\,B_2)\;
\big( A_1\; (-)^{\sum_{k=1}^3 s_k}\;\epsilon^{(+)}_{(s_1s_2s_3 0)}
+ A_2\; (-)^{\sum_{k=2}^3 s_k}\;\epsilon^{(+)}_{(\bar s_1 s_2s_3 0)}\big)\cr
\Gamma_{\ob}\;\eta^{(\pm)}_{(s_1s_2s_3)} &=& \pm\;\eta^{(\pm)}_{(s_1s_2s_3)}\label{kill42}
\eea
It is easy to see that the $\bar D 4$-brane solution (corresponding to have
a minus sign in the ${\ob}\,$-condition) will have the same Killing spinors provided that it
has {\it both } electric and magnetic fields with opposite signs wrt that of the $D4$-brane.
Therefore (\ref{kill2}) with $\eta^{(+)}_{(s_1s_2s_3)}$ given in (\ref{kill42}) is the
general Killing spinor of such brane-antibrane systems.

It is worth to note however that (\ref{const42}) implies
$0<|| B^{-1}\vec E||^2 = 1+ \left(\frac{E_1}{B_1}\right)^2 +
\left(\frac{E_2}{B_2}\right)^2 =1- \frac{1}{B_1{}^2\,B_2{}^2} <1\;\;, $
and therefore from the footnote at the beginning of this Section we know that there exists
a boost with $\vec\beta = - B^{-1}\vec E\,$
that eliminates the electric field; a further rotation will lead us to the case
$\vec E = \vec 0$ (with {\it different} $B_1,\,B_2$).
The Majorana condition in this case looks like in (\ref{majo41}).

\section{Conclusions}
\cleqn

We have studied in the context of the Born-Infeld effective action the existence of
supersymmetric, presumably  stable solutions of $D2$, $D3$ and $D4$-branes  preserving a
quarter of the supersymmetries of the flat background in which they are embedded.
These results allow to conjecture at the light of the compatible ansatz in solutions II, the existence of
configurations other than the T-dual to those presented here, for $D5$ and $D6$ branes , and from here for
$D7-\bar D7$ and $D8-\bar D8$ and so on.
An explicit prove of theses facts along the lines followed here should be straightforward.

In the case of the $D4$ brane we note that equations (\ref{conda42}), (\ref{condb42}) imply the existence of
Taub-NUT charge and a sort of $D 2$-brane charge for this solution, no $D4$-brane charge is present;
the $D 4$-$\bar D 4$ systems should represent genuine bound states of these components.
What is more, it is plausible that a five dimensional supertube-like solution exists,
leading in a certain limit to the brane-antibrane system much as it happens with the supertube.

Another interesting open problem is certainly to find the explicit form of the corresponding
supergravity solutions since as it is stressed in reference \cite{sutubo2} the low energy analysis presented
here does not assure by itself the complete absence of instabilities.
To this goal the knowledge of the world-volume fields as well as the explicit form of the Killing spinors
(although they do not take into account the back-reaction) could be of great help.
All these items are under current investigation \cite{lr}.

\section*{Acknowledgements}
\cleqn

We would like to thank to Jos\'{e} Edelstein for discussions and specially Jorge Russo for
discussions and comments on the manuscript.

\appendix\section{Appendix}
\cleqn

We briefly summarize conventions and formalism used in the text
(see for example \cite{wein} ).

We start with the representation of the anti-commutation relation $\{b;b^\dagger\}=1$
in a two-dimensional space expanded by the vectors $|s>$ where $s=1$ (spin up) or $s=0$
(spin down).
In this basis we take
$|1>\rightarrow (1)\equiv\left(\begin{array}{l}1\\0\end{array}\right)\;$,
$|0>\rightarrow (0)\equiv\left(\begin{array}{l}0\\1\end{array}\right)\;$.
Then the representation of the fermion algebra is given by the matrices
\be
b =\left(\begin{array}{ll}0&0\\1&0\end{array}\right)\;\;\;\;\;,\;\;\;\;\;\;
b^\dagger = \left(\begin{array}{ll}0&1\\0&0\end{array}\right)
\ee
Then $\Gamma_1\equiv b + b^\dagger = \sigma_1$ and
$\Gamma_2\equiv -i\,(b - b^\dagger) = -\sigma_2\;$ satisfy
$\{\Gamma_M;\Gamma_N\}=2\,\delta_{MN}\;,\, M,N=1,2\;$,
where $\{\sigma_i, i=1,2,3\}$ are the Pauli matrices.
Thus $\sigma_1(s) = (\bar s)\;,\; \sigma_2(s) = i(-)^{\bar s}(\bar s)\;,\;
\sigma_3(s) = (-)^{\bar s}(s)\;$, where we define $\bar s\equiv 1-s$.
The number operator
$N\equiv b^\dagger b$
obeys $N|s> = s |s>$.

The Weyl basis for the vector space of spinors in $d=10$ can be constructed as the
tensor product of five copies of the space defined above, and consists therefore of
$32$ vectors denoted by $(s_1)\otimes\dots\otimes(s_5)\equiv(s_1\dots s_5)$.
The euclidean Clifford algebra $\{\Gamma^M;\Gamma^N\}=2\,\delta_{MN}$ is realized in
this space by the matrices
\bea
\Gamma_{2k-1} &=& (-)^{k-1}\;\sigma_3\otimes\dots\otimes\sigma_3\otimes\sigma_1\otimes 1\otimes\dots\otimes 1\cr
\Gamma_{2k} &=& (-)^{k}\;\;\;\;\,\sigma_3\otimes\dots\otimes\sigma_3\otimes\sigma_2\otimes 1\otimes\dots\otimes 1
\eea
where $k=1,\dots,5$, and $\sigma_1$, respectively $\sigma_2$, is placed in the
$k$-position.
We will omit in general the tensor product symbol $\otimes$.
By definition,
$\Gamma_0 \equiv i \Gamma_{10}=
-i\,\sigma_3\,\sigma_3\,\sigma_3\,\sigma_3\,\sigma_2\,$,
and then $\{\Gamma_M;\Gamma_N\}=2\,\eta_{MN},\, M,N=0,1,\dots 9\,$, where
$(\eta_{MN})= \left(\matrix{-1&0\cr 0&1_9}\right)$, $1_n$ standing for the
$n\times n$ identity matrix.

The chirality matrix is defined by
\bea
\Gamma_{11} &\equiv& i\;\Gamma_1\dots\Gamma_{10} = \Gamma_1\dots\Gamma_9\;\Gamma_0 =
\sigma_3\,\sigma_3\,\sigma_3\,\sigma_3\,\sigma_3\cr
\Gamma_{11}\, (s_1\dots s_5) &=& (-)^{\sum_{k=1}^5 \bar s_k}\;(s_1\dots s_5)
\eea
from where it follows that states with number of spins down even (odd) have positive
(negative) chirality.

As usual we denote $\Gamma_{M_1\dots M_n}, n\leq 10$, to the completely
antisymmetric product of the $n\, \Gamma$-matrices, i.e.
$\Gamma_{1\dots n} = \Gamma_1 \dots \Gamma_n$.
In particular $S_{MN}\equiv S(X_{MN})= \frac{1}{2}\Gamma_{MN}$ gives the spinorial
representation of the Lorentz generators satisfying the standard algebra
\be
[X_{MN}, X_{PQ}] = \eta_{MQ}\;X_{NP} + \eta_{NP}\;X_{MQ} - (M\leftrightarrow N)
\label{la}
\ee
However this representation is reducible due to the fact that $[\Gamma_{11};S_{MN}]=0$.
This leads to define $S^{(\pm)}\equiv \frac{1}{2}\,(1\pm\Gamma_{11})\,S_{MN}$; thus
$S_{MN} = S^{(+)}_{MN}+ S^{(-)}_{MN}$ decomposes in two irreducible representations
with $\Gamma_{11}=+1$ (positive chirality) and $\Gamma_{11}=-1$ (negative chirality)
of the Lorentz algebra, or Weyl-left and Weyl-right spinors space respectively.

A Majorana condition is a reality condition; to be able to define it in some fixed basis
we need a matrix $D$ satisfying $D^{-1}\, S_{MN}\, D = S_{MN}^*$ in such a way that if
$\Psi$ is a spinor, e.g. a $32$-dimensional vector (in $d=10$) transforming linearly in the
spinorial representation of the Lorentz group, then $\Psi^c\equiv D\,\Psi^*\,$ also it
is; $\Psi^c$ is called the conjugate spinor \cite{son}.
Then it has sense to impose the Majorana condition
\be
\Psi^c \equiv D\;\Psi^* = \Psi\label{majo}
\ee
In the Weyl basis a matrix $D$ verifying
$D^{-1}\, \Gamma_M\, D  =\Gamma_M^*\;,M,N,=0,1,\dots,9\,$, can be taken as
\be
D \equiv -\Gamma_2\;\Gamma_4\; \Gamma_6\;\Gamma_8 = \sigma_1\;\sigma_2\;\sigma_1\;\sigma_2\; 1
\ee
It is worth to note that if $D$ is such a matrix in a basis $\{|\alpha>, \alpha=1,\dots,32\}$,
under a change of basis $|\alpha> = P^\beta{}_\alpha\,|\beta>'$ it transforms as
$D'=P^{-1}\,D\,P^*$.


\end{document}